\documentclass[twocolumn,prl,showpacs]{revtex4}
\usepackage{graphicx}

\begin{document}

\title{Strain enhancement of superconductivity in CePd$_2$Si$_2$ under pressure}

\author{A. Demuer}
\author{A.~T. Holmes}
\author{D. Jaccard}
\affiliation{DPMC, University of Geneva, 24 quai Ernest Ansermet, 1211 Geneva 4, Switzerland.}
\begin{abstract}
We report resistivity and calorimetric measurements on two single crystals of CePd$_2$Si$_2$ pressurized up to
7.4$\:$GPa. A weak uniaxial stress induced in the pressure cell demonstrates the sensitivity of the physics to
anisotropy. Stress applied along the $c$-axis extends the whole phase diagram to higher pressures and enhances the
superconducting phase emerging around the magnetic instability, with a 40\% increase of the maximum superconducting
temperature, $T_c$, and a doubled pressure range. Calorimetric measurements demonstrate the bulk nature of the
superconductivity.

\end{abstract}
\maketitle

By varying an external control parameter, such as magnetic field, composition or pressure, many heavy-fermion systems
may be pushed through a quantum critical point (QCP), where their magnetic ordering temperature goes to zero. In the
immediate vicinity of this point, transport and thermodynamic measurements show striking deviations from standard
Fermi-liquid behavior~\cite{Lohneysen94,Kambe96,Braithwaite98,Cambridge,Sheikin01,Demuer01,Raymond99}. In particular,
the low temperature resistivity, $\rho(T)$, exhibits a $T^n$ behavior with 1$<$$n$$\leq$1.5 over a wide temperature
range. The nature of this non-Fermi-liquid (NFL) behavior remains an open question~\cite{Schroder00}. Is the
spin-fluctuation description~\cite{Kambe96} appropriate, with itinerant magnetism developing below a characteristic
temperature such as the Kondo temperature, $T_K$? Alternatively, this characteristic temperature may collapse at the
QCP, leading to localized magnetism down to the lowest temperatures~\cite{Lohneysen94,Schroder00}. Particular attention
has been paid to the case of stoichiometric compounds, whose weak disorder permits the observation of superconductivity
around the magnetic instability. Due to the enhancement of low-lying magnetic excitations in this region, it is
commonly believed that Cooper pairs are magnetically formed. In one of these systems, CePd$_2$Si$_2$, superconductivity
was discovered in a window of about 1$\:$GPa around the QCP at a critical pressure $P_c\simeq2.8\:$GPa. Simultaneously,
NFL behavior was found in resistivity measurements with a $T^{1.2-1.3}$ law over two decades in
temperature~\cite{Cambridge,Sheikin01,Demuer01,Raymond99}. As an exponent $d/2$ is predicted for a {\it d}-dimensional
antiferromagnet by spin-fluctuation theory~\cite{Millis,Moriya95,Lonzarich}, it has been suggested that the magnetic
excitation spectrum has an effective dimension close to 2. This assumption is supported by the quasi-linear pressure
dependence of the N\'{e}el temperature, $T_N$, predicted to be $(P_c-P)^{2/d}$, by the tetragonal symmetry ($I$4$/mmm$)
and by the magnetic structure containing a frustrated moment in the center of the elementary cell~\cite{Cambridge}.

The results quoted above were obtained in hydrostatic conditions using a ``liquid'' pressure transmitting medium.
Another investigation~\cite{Raymond99} was carried out in a Bridgman anvil cell, using a soft solid (steatite) as a
pressure transmitting medium~\cite{Eremets96}. This showed a rather different phase diagram: around a higher critical
pressure $P_c\simeq3.6\:$GPa, a strikingly expanded superconducting region was found, lying from 2 to 7 GPa with a
maximum of $T_c$ apparently disconnected from $P_c$, casting doubt upon spin fluctuations as the only mediation
mechanism for superconductivity. As this pressure technique is suspected to provide higher pressure gradients, a
residual stress along the cell axis could be at the origin of these differences. Our motivation was thus to demonstrate
and understand the effect of uniaxial stress under pressure.

In this letter, we report resistivity and calorimetric measurements performed on two samples in a Bridgman anvil cell
up to 7.4$\:$GPa. The samples were set in the pressure cell with the force load direction perpendicular and parallel to
the $c$-axis (ref.~\cite{Raymond99} being the latter). These measurements demonstrate the high sensitivity of the
physics in CePd$_2$Si$_2$ to pressure conditions, and the crucial influence of anisotropy on the emerging
superconductivity. The differences between the previously observed phase diagrams can be explained by the results from
these two samples, with an enhancement of superconductivity when uniaxial stress is applied along the $c$-axis.
Calorimetric measurements demonstrate the bulk nature of this superconductivity; a combination of both types of
measurement leads to further insight into the quantum critical point and its associated energy scale.

\begin{figure}[t]
\includegraphics{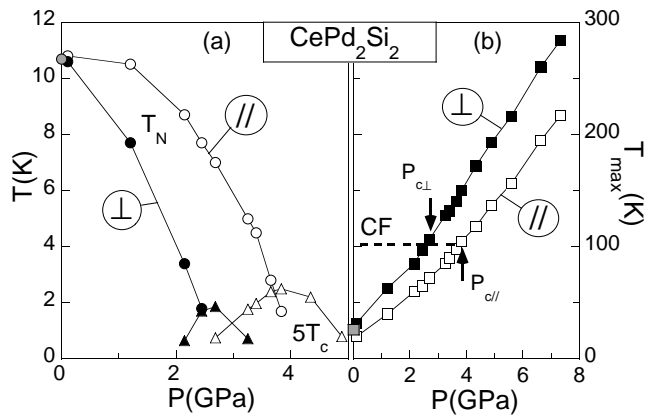} \caption{(a) Phase diagram of the two samples (filled and open symbols for samples
$\perp$ and $/\!/$ respectively) pressurized as described in the text.(b) Pressure dependence of $T_{max}$, the
temperature of the maximum in the magnetic part of $\rho(T)$. The dashed line, which qualitatively indicates the
position of the crystal-field (CF) contribution, crosses $T_{max}$ close to $P_c$ for both samples. Gray symbols
indicate values at $P=0\:$GPa.} \label{figure1}
\end{figure}

The samples were extracted from the same single crystalline platelet used in refs.~\cite{Sheikin01,Demuer01,Raymond99}.
A parallepiped sample (510$\times$75$\times$60$\:\mu$m$^3$), with a residual resistivity ratio $RRR\!\simeq\:$62, was
cut into two pieces of length 250$\:\mu$m. These were polished to a small cross-section
$(\sim$70$\times$20$\:\mu$m$^2$), and spot-welded with $5\:\mu$m diameter gold wires, giving
$\rho(293\:$K)$=45\:\mu\Omega\;$cm to within 10\% for both samples and $RRR$ values of 48 and 103. The corresponding
residual resistivities, $\rho_0$, were respectively 1 and $0.48\:\mu\Omega\;$cm. The samples will be referred to as
$/\!/$(higher $\rho_0$) and $\perp$ (lower $\rho_0$) in relation to the orientation of their $c$-axis with respect to
the force load direction (and the additive uniaxial stress). Both samples were connected for four-point {\small DC}
resistivity measurements, with sample $\perp$ having additional connections for a constantan resistive heater and a
thermocouple Au/Au-0.07 at.\% Fe suitable for {\small AC} calorimetric measurements \cite{Bouquet00}. The pressure was
determined by the superconducting transition of a lead manometer.

Sample $\perp$ gave rise to a phase diagram similar to that obtained in hydrostatic conditions (Fig.~1). The
superconductivity was limited to the range 2.14-3.25$\:$GPa around $P_{c\perp}\simeq2.7\:$GPa, with $T_{c\perp}$ having
a maximum of 375$\:$mK (mid-point criterion). In contrast, the phase diagram of sample $/\!/$ seems to be stretched
towards higher pressures. $T_N$ collapses at $P_{c/\!/}=3.9\:$GPa with a critical behavior $(P-P_c)^\alpha$,
$\alpha=0.60\pm0.05$, as distinct from the quasi-linear dependence in the hydrostatic case. Superconductivity occured
between 2.14 and 5.0$\:$GPa (using a mid-point criterion). As in the previous investigation in a Bridgman
cell~\cite{Raymond99}, $T_c$ reached a higher value, 520$\:$mK in our case. This maximum of $T_c$ coincides with $P_c$,
suggesting that this extended superconductivity is still related to the QCP. The apparent discrepancy between $P_c$ and
the maximum of $T_c$ in ref.~\cite{Raymond99} can be explained by the criterion chosen (onset), sensitive to the large
transition widths at extremes pressures.
\begin{figure}[!t]

\includegraphics{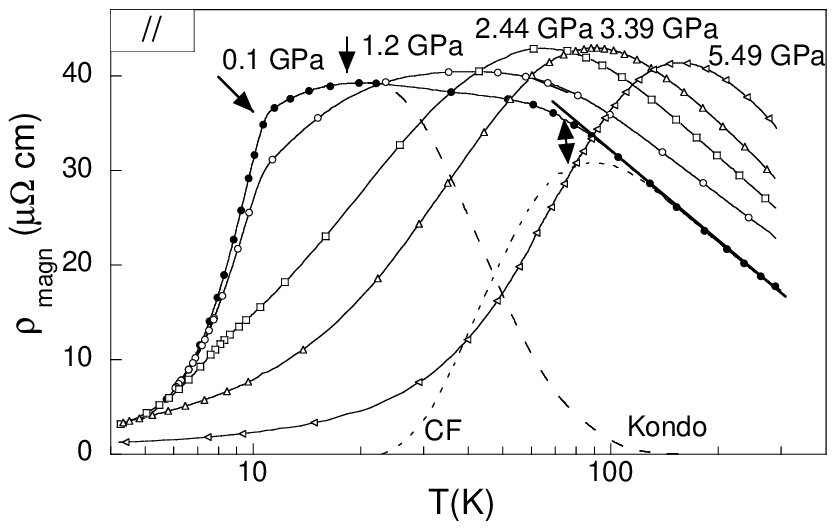} \caption{Magnetic contribution to $\rho(T)$ of sample $/\!/$ at selected pressures after
substraction of a linear term assumed for phonons. On the $P=0.1\:$GPa curve, the arrows indicate $T_N$, the Kondo peak
at $T_{max}$ and a shoulder attributed to the two excited crystal-field (CF) levels. The solid line shows the $-\ln T$
Kondo dependence at high temperature. Dashed lines qualitatively represent the Kondo and CF contributions.}
\label{figure2}
\end{figure}

Fig.~2 shows $\rho(T)$ curves from sample $/\!/$ for selected pressures with a phononic linear contribution
($0.1\:T\:\mu\Omega\;$cm/K) subtracted. The first pressure, 0.1$\:$GPa, corresponds mainly to a small uniaxial stress
along the force load direction. With increasing temperature, one can distinguish a clear kink at $T_N\simeq11\:$K, a
maximum at $T_{max}$ attributed to the Kondo effect and a ``shoulder'' reflecting the influence of excited
crystal-field (CF) levels. At high temperature, the $-\ln T$ dependence is characteristic of Kondo scattering. As the
pressure rises, $T_{max}$ increases continuously whereas the excited CF anomaly is rather pressure independent. This
latter progressively merges with the Kondo peak around 1.5$\:$GPa and seems to collapse at higher pressures.
$T_{max}(P)$ shows no anomaly at $P_c$ and identical values $T_{max}(P_c)$ in both samples (Fig.~1).

The resistivity was analyzed at low temperature in terms of a power law $\rho (T)=\rho_0+AT^n$. Such dependencies are
not stable over a wide temperature range except for pressures close to $P_c$, where these power laws extend up to
30$\:$K. The fits were therefore limited to a window of 0.5-2$\:$K, in order to compare data over the entire pressure
range. Fig.~3 shows the pressure dependence of the coefficient $A$ and the exponent $n$ (inset). $A$ behaves as
$(d\rho/dT^2)_{T\rightarrow 0}$ and may be interpreted as a Fermi-liquid contribution prefactor. As expected from the
spin-fluctuation model SCR~\cite{Moriya95}, $A(P)$ shows a sharp maximum at $P_c$. This maximum is similar in both
samples with $A(P_c)/A(0)\simeq5$. At 7.4$\:$GPa, the $A$ coefficient of sample $\perp$ has fallen by a factor of 100
compared to its value at $P_c$. A small anomaly was found in $A(P)$ at about 1$\:$GPa in sample $\perp$, and 2$\:$GPa
in sample $/\!/$, possibly corresponding to a pressure induced magnetic phase transition. Such an anomaly seems to be
present in the isoelectronic compound CePd$_2$Ge$_2$ at about 12$\:$GPa just below
$P_c\simeq13.8\:$GPa~\cite{Wilhelm02}.

An additional curve in Fig.~3 shows the pressure dependence of $\gamma^2=(C/T)^2_{T\rightarrow 0}$ in sample $\perp$;
this also has a maximum at $P_c$. For each pressure, $\gamma^2$ was estimated at 100$\:$mK by subtraction of 1/$V^2$
taken at two frequencies (16 and 256$\:$Hz) where $V$ is the thermocouple voltage amplitude~\cite{Bouquet00}. To obtain
a reliable pressure dependence, the same working parameters were used for all pressures. Far from the instability, $A$
and $\gamma$, both related to the square of the effective mass of quasi-particles, $m^{\ast 2}$, are expected to follow
the Kadowaki-Woods relation, $A\propto\gamma^2$~\cite{Kadowaki86}. The peak in $\gamma^2$ at $P_c$ is less pronounced
than in $A(P)$ with $\gamma^2(P_c)/\gamma^2(7.4\:$GPa)$\simeq18$, but $\gamma$ may well include a contribution from the
pressure transmitting medium, reducing the relative size of the peak.

In both samples, a sharp dip in the resistivity exponent $n(P)$ (inset of Fig.~3) is associated with the magnetic
instability, reaching values lower than the 2 expected for Fermi-liquid behavior. The minimum values obtained were 1.32
and 1.42 ($\pm$0.03) for samples $\perp$ and $/\!/$ respectively. As in the $A(P)$ curves, a small anomaly appears
around 1 and 2$\:$GPa.

\begin{figure}[!t]
\includegraphics{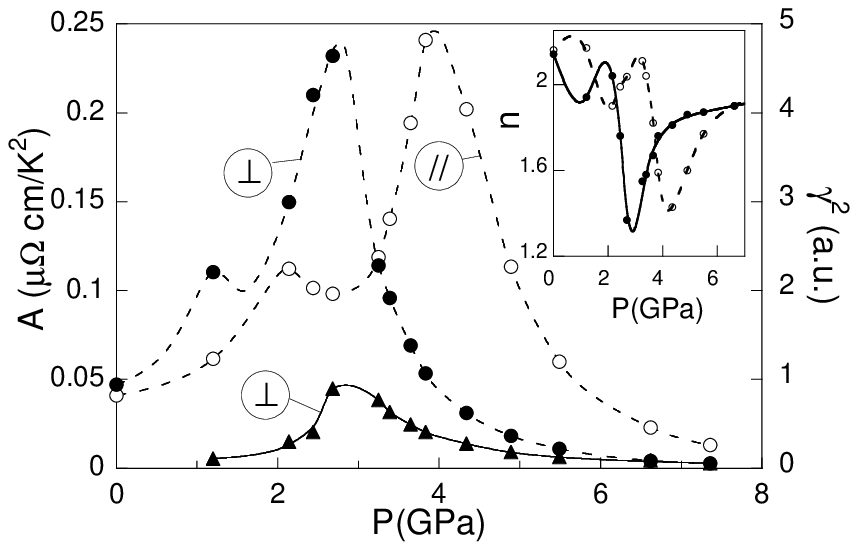} \caption{$A$ coefficient (circles) and temperature exponent $n$ (inset) of the
low-temperature resistivity ${\rho=~\rho_0+AT^n}$ for both samples. Open and filled symbols indicate $/\!/$ and $\perp$
samples respectively. Filled triangles show $\gamma^2$, estimated for sample $\perp$ (see text), normalized at
7.4$\:$GPa to the value of $A$ at the same pressure.  Lines are guides for the eye.} \label{figure3}
\end{figure}

The superconducting transition appears in the calorimetric measurement only at 2.68$\:$GPa, the closest pressure to
$P_c$. Fig.~4 shows a comparison between superconducting transitions in $\rho(T)$ and the calorimetric signal 1/$V$
($\sim C_p/T$). The onset of the calorimetric transition occurs at the temperature for which $\rho$ reaches zero. In
sample $\perp$ at 2.68$\:$GPa$\:\simeq P_c$, we studied the effect of an external magnetic field on the
superconductivity using the two types of measurement. The large initial slope of the upper critical magnetic field in
the basal plane, $dH_c^a/dT\simeq-6\:$T/K, indicates that heavy quasi-particles are involved in superconductivity. The
size of the calorimetric anomaly collapses rapidly with increasing field and becomes undetectable above 0.5$\:$T. If
the calorimetric anomaly at $H$=0 indicates a bulk transition, one cannot rule out the magnetic field revealing a
non-homogeneous situation in the sample, as suggested by the large transition widths in $\rho(T)$ and disappearance of
the anomaly in calorimetric measurement for pressures away from $P_c$. The inset in Fig.~4 shows the superconducting
transition in $\rho(T)$ for sample $/\!/$ close to its $P_c$. A striking point is that the highest value of $T_c$ is
obtained in the sample with the larger residual resistivity, demonstrating that the superconductivity enhancement is
not related to the crystal purity.

\begin{figure}[!t]
\includegraphics{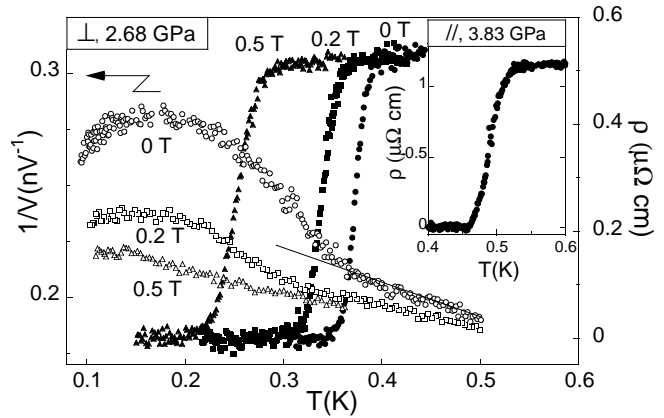} \caption{Effect of magnetic field on resistive and calorimetric
superconducting transitions (filled and open symbols) for sample $\perp$ close to $P_c$. The inset shows the
superconducting transition in $\rho(T)$ of sample $/\!/$ close its $P_c$.} \label{figure4}
\end{figure}

The following discussion is supported by the unprecedented quality of our samples, with $RRR$ values as high as 130 at
$P_c$. Bearing this in mind, one should be aware that variations on a submillimetric scale exist even within a single
crystal - the present samples and those of ref.~\cite{Sheikin01,Demuer01} were cut from the same tiny
single-crystalline platelet with $RRR$ values varying by a factor of 3 at $P=0$. Furthermore, we claim to have an
accurate value for the resistivity, with a well-defined geometric factor enabling the determination of the absolute
resistivity to within 10\%. As suggested earlier, the temperature $T_{max}$ of the maximum in the magnetic contribution
to the resistivity should be related to the Kondo temperature, $T_K$. $T_{max}$ takes the same value at $P_c$ for both
samples, supporting the idea that $T_{max}$ is a reliable characteristic energy for the QCP. Its pressure dependence
allows us to take part in the heated debate about the nature of the QCP illustrated by another heavy-fermion system,
CeCu$_{6-x}$Au$_x$. While neutron measurements on the substituted compound CeCu$_{5.9}$Au$_{0.1}$ seem to reveal
localized magnetism at the lowest temperature~\cite{Lohneysen94}, measurements under pressure on the stoichiometric
compound CeCu$_5$Au~\cite{Wilhelm00} showed no anomaly in $T_K$ at $P_c$, suggesting that the magnetism remains
itinerant around the QCP. As the latter behavior is observed in our investigation, spin-fluctuation theory should also
apply in the vicinity of the QCP of CePd$_2$Si$_2$. In both of our samples, the QCP occurred in a pressure domain where
the characteristic Kondo energy $k_BT_K$ ($T_K\propto T_{max}$) typically reaches the crystal-field splitting energy
(see Fig.~1). Furthermore, $\ln A$ is found to behave as $-\alpha\ln T_{max}$ with a slope $\alpha\simeq4$, instead of
the value of 2 expected for a normal heavy-fermion regime. This indicates the entrance into an intermediate valence
regime, probably leading to deviations from the simple spin-fluctuation model. The fact that $\gamma^2(P)$ decreases
slower than $A(P)$ above $P_c$ might tempt us to invoke the predictions of ref.~\cite{Takimoto96}, but our calorimetric
measurement is not quantitative enough. It does allows us to demonstrate clearly a relationship between $A$ and
$\gamma$, showing in both a peak at $P_c$, but the $\gamma$ value extracted probably includes undefined addenda
obscuring the physics.

As in previous measurements, NFL behavior was observed in $\rho(T)$ at $P_c$ in both samples over more than one decade
in temperature. The stability of this behavior in temperature has been proposed to result from a crossover between
``clean'' and ``dirty'' limit regimes for a specific amount of disorder ~\cite{Rosch99}. However, this explanation
disagrees with the systematic observation of power laws in $\rho(T)$ at $P_c$ over a large temperature range for
samples with residual resistivities spread over almost one decade~\cite{Cambridge,Sheikin01,Demuer01,Raymond99}. The
exponent in $\rho(T)=\rho_0+AT^n$ in all cases reaches remarkably low values with $n\simeq1.2-1.3$, a value generally
attributed to a non 3D spin-fluctuation spectrum. However, let us recall that in other compounds such as
CeCu$_2$Ge$_2$, a minimum of $n$ close to 1 was found only for $P>P_c$, in a pressure domain where $k_BT_K$ reaches the
CF splitting energy~\cite{Jaccard99}. As this happens for $P\simeq P_c$ in CePd$_2$Si$_2$, one may wonder if the low
exponent observed is not a consequence of this change of regime.

Hydrostatic pressure reduces both lattice parameters $a$ and $c$. At a given pressure, an additional uniaxial stress,
$\sigma$, along one axis further reduces that lattice parameter while expanding the others. A description of the
physical properties as a function only of the cell volume fails in this system, as shown by the various phase diagrams
obtained on samples $\perp$ and $/\!/$. Whereas the situation remains mostly unchanged when $\sigma$ is applied in the
basal plane, the clear extension of the phase diagram for $\sigma$ along the $c$-axis shows that the ratio $c/a$,
reflecting the anisotropy of the system, is also a key parameter. Considering the spin-fluctuation prediction
$T_N(P)\propto(P_c-P)^{2/d}$, the exponent $0.60\pm0.05$ obtained for sample $/\!/$ suggests that applying $\sigma$
along the $c$-axis restores a 3D spin-fluctuation spectrum. With the same theoretical approach, the minimum value of
the exponent in the $\rho(T)$ power law at $P_c$, predicted to be  $T^{d/2}$, should be different in the two samples.
However, the differences observed in $n(P_c)$ and in $A(P_c)$ (Fig.~3) are smaller than we might expect. The most
striking consequence of this change in anisotropy is the apparent enhancement of superconductivity for $\sigma$ applied
along the $c$-axis with a 40\% higher maximum value of $T_c$ and a doubling of its pressure range. This enhancement is
not related to a larger electronic mean free path due to reduced disorder, since the sample with a higher $T_c$ also
has the higher $\rho_0$. As spin fluctuations are thought to be at the origin of the Cooper pairing, one may attribute
the enhancement of superconductivity to different features of the spin-fluctuation spectrum. The values of the critical
exponents in $T_N(P)$ suggests that 3D spin fluctuations would be more favorable for superconductivity, though many
scenarios such as an increase in carriers density associated with a band modification under uniaxial stress remain
possible.

Our measurements demonstrate the complexity of the physics in CePd$_2$Si$_2$ in the vicinity of its quantum critical
point. At this pressure $P_c\simeq2.7-2.8\:$GPa, several energy scales such as the Kondo and excited crystal-field
energies interact, leading to a complex ground state. While the other archetypical system for a superconducting phase
induced around its critical point, the cubic CeIn$_3$, is insensitive to pressure conditions~\cite{Cambridge,Knebel01},
the physical properties of the tetragonal CePd$_2$Si$_2$ are strongly affected by modification of anisotropy resulting
from additional uniaxial strain along the $c$-axis. The quasi 2D-behavior evoked for spin-fluctuations seems to be
destroyed and superconductivity is enhanced around $P_c\sim3.9\:$GPa. As pure uniaxial stress experiments are extremely
difficult to perform under pressure, the effect of the anisotropy on superconductivity around a quantum critical point
should be checked on a compound close to its instability at ambient pressure. CeNi$_2$Ge$_2$, where traces of
superconductivity as well as quasi-2D behavior for spin fluctuations were found~\cite{Braithwaite98,Fak00}, appears as
one of the best candidates.

\newpage


\begin{thebibliography}{99}
\bibitem{Kambe96}S. Kambe {et al.}, Physica B {\bf 223 \& 224}, 135 (1996)
\bibitem{Lohneysen94}H.v. L\"{o}hneysen {\it et al.}, \prl {\bf 72}, 3262 (1994)
\bibitem{Braithwaite98}D. Braithwaite {\it et al.}, J. Phys.: Condens. Matter {\bf 12}, 1339 (2000)
\bibitem{Cambridge}N.D. Mathur {\it et al.}, Nature {\bf 394}, 39 (1998)
\\ F.M. Grosche {\it et al.}, J. Phys.: Condens. Matter {\bf 13}, 2845 (2001)
\bibitem{Sheikin01}I. Sheikin {\it et al.}, J. Low Temp. Phys. {\bf 122}, 591 (2001)
\bibitem{Demuer01}A. Demuer {\it et al.}, J. Phys.: Condens. Matter {\bf 13}, 9335 (2001)
\bibitem{Raymond99}S. Raymond and D. Jaccard, \prb {\bf 61}, 8679 (2000)
\bibitem{Schroder00}A. Schr\"{o}der {\it et al.}, Nature {\bf 407}, 351 (2000)
\bibitem{Millis}A.J. Millis, \prb {\bf 48}, 7183 (1993)
\bibitem{Moriya95}T. Moriya and T. Takimoto, J. Phys. Soc. Jpn {\bf 64}, 960 (1995)
\bibitem{Lonzarich} G.G. Lonzarich in {\it Electron: A Centenary Volume} edited by M. Springford (Cambridge University
Press, 1999)
\bibitem{Eremets96}M. Eremets, {\it High Pressure Experimental Methods} (Oxford University, Oxford, 1996)
\bibitem{Bouquet00}F. Bouquet {\it et al.}, Solid. State Commun. {\bf 113}, 367 (2000)
\bibitem{Wilhelm02}H. Wilhelm and D. Jaccard, to be published.
\bibitem{Kadowaki86}K. Kadowaki and S.B.  Woods, Solid. State Commun. {\bf 58}, 507 (1986)
\bibitem{Takimoto96}T. Takimoto and T. Moriya, Solid State Commun. {\bf 99}, 457 (1996)
\bibitem{Wilhelm00}H. Wilhelm {\it et al.}, {\it in Sci. and Technol. High Pressure} edited by M.H. Manghnani, W.J. Nellis and
M.F. Nicol (Universities Press), 697 (2000); also cond-mat/9908442
\bibitem{Rosch99}A. Rosch, \prl {\bf 82}, 4280 (1999)
\bibitem{Jaccard99}D. Jaccard {\it et al.}, Physica B {\bf 259-261}, 1 (1999)
\bibitem{Knebel01}G. Knebel {\it et al.}, \prb {\bf 65}, 024425 (2002)
\bibitem{Fak00}B. F\aa k {\it et al.}, J. Phys.: Condens. Matter {\bf 12}, 5423 (2000)
\end{thebibliography}
\end{document}